\documentclass[a4paper]{article}
 
\usepackage{array}

\usepackage{graphicx}

\usepackage{tabularx}

\usepackage{enumerate}

\usepackage[numbers]{natbib}
\usepackage{amsmath}
\usepackage{amssymb}
\newcommand{\txt}[1]{\quad\text{#1}\quad}

\newcommand{\sech}{\operatorname{sech}}

\newcommand{\half}{\frac{1}{2}}

\begin{document}

\title{Soliton formation from a pulse passing
the zero-dispersion point in a nonlinear Schr\"odinger equation}
\author{S.R. Clarke, R.H.J. Grimshaw\\
Department of Mathematics and Statistics\\
Monash University, Clayton, Victoria 3168, Australia\\
\noalign{\medskip}
and \\
\noalign{\medskip}
Boris A. Malomed\\
Department of Interdisciplinary Studies, Faculty of 
Engineering\\
Tel Aviv University, Tel Aviv 69978, Israel}

\maketitle

\begin{abstract}

We consider in detail the self-trapping of a soliton
from a wave pulse that passes from a defocussing region into a
focussing one in a spatially inhomogeneous nonlinear waveguide,
described by a  nonlinear Schr\"odinger equation
in which the dispersion coefficient changes its sign from normal
to anomalous. The model has direct applications to
{\it dispersion-decreasing} nonlinear
optical fibers, and to natural waveguides for internal waves in the
ocean. It is found that, depending on the (conserved) energy and 
(nonconserved) ``mass" 
of the initial pulse, four qualitatively different outcomes of
the pulse transformation are possible: decay into radiation; self-trapping
into a single soliton; formation of a breather; and formation of a
pair of counterpropagating solitons. A corresponding chart is drawn on
a parametric plane, which demonstrates some
unexpected features. In particular, it is found that any kind of
soliton(s) (including the breather and counterpropagating pair) 
eventually decays into pure radiation
with the increase of the energy, the initial ``mass" being kept constant. 
It is also noteworthy that a virtually direct transition from
a single soliton into a pair of symmetric counterpropagating ones seems
possible. An explanation for these features is proposed. In two
cases when analytical approximations apply, viz., a simple
perturbation theory for broad initial pulses, or the variational
approximation for narrow ones, comparison with the direct simulations
shows reasonable agreement.

\end{abstract}

\newpage

\section{Introduction}

Nonlinear spatially inhomogeneous waveguides give rise to a number
of effects which are of interest by themselves, and also find important 
applications in such diverse fields as optical {\it dispersion-decreasing
fibers} (DDF) \cite{boga91} and natural waveguides for internal waves in
the ocean with a shear-flow background
\cite{grim97}. It is easy to understand that most nontrivial
effects take place in the vicinity of a critical point, where the waveguide's
dispersion or nonlinear coefficient changes sign.

The critical points corresponding to the wave propagation of the
nonlinear-Schr\"{o}dinger (NLS) type were classified earlier in
Ref. \cite{malo91}, where it was demonstrated that the most 
interesting one is that at which the sign of the dispersion coefficient
$\alpha$ 
changes. However, unlike the case when the nonlinear coefficient 
in the corresponding NLS equation changes its sign, this case
is not amenable to a consistent analytical consideration, hence
systematic numerical simulations are necessary. 

In a very recent paper \citep{ccgm99}, the propagation of a wave pulse
in this model was simulated for a situation when $\alpha$ changes its sign,
in a self-focussing medium, from anomalous (admitting the existence 
of bright solitons) to normal (for which bright solitons do not exist).
Accordingly, disintegration of an initial
soliton-like pulse into radiation wave fields was considered. Despite 
the apparent simplicity of the process, a number of quite nontrivial
features were found and qualitatively explained, the most interesting one 
being a double-humped structure in the region of the normal
dispersion. 

For applications, particularly for those related to nonlinear 
optics of fibers and planar waveguides, 
especially relevant is the reverse process, i.e., formation of a soliton 
from a wave pulse
crossing into the anomalous-dispersion region from the normal-dispersion
one. Previously,
this process was considered by one of the present authors in
\cite{malo93} in a purely analytical approximation, based on 
a variational technique. As we will demonstrate in this work,
the system of ordinary differential equations derived in Ref. 
\cite{malo93} from the 
underlying variable-coefficient NLS equation by means of the
variational approximation, indeed provides for quite an accurate
description of the pulse's dynamics in a parametric region where
the approximation is relevant. Nevertheless, most results to be
reported in the present work were produced by systematic 
direct simulations of the NLS equation.

A modified NLS equation, valid 
near the zero-dispersion point for the
nonuniformly guided nonlinear wave propagation,
was introduced in the previous works \cite{malo93,ccgm99}:
\begin{equation}
iu_z + \alpha(z) u_{tt} + 2
|u|^2u = i\delta u_{ttt},\label{mnls}
\end{equation}
where $u$ is the local amplitude of the guided wave,
$z$ and $t$ are the distance along the waveguide
and the so-called reduced time (see, e.g., the derivation
of the corresponding nonlinear Schr\"{o}dinger 
equation for optical fibers in the book \cite{agra95}),
$\alpha(z)$ is the above-mentioned sign-changing variable 
dispersion coefficient, and $\delta$
is the third-order-dispersion (TOD) coefficient
which, generally, should be included in the
case when the usual dispersion becomes very weak \cite{agra95}. 

Here we consider solutions to Eq. \eqref{mnls} for a model with 
a continuous piecewise-linear dispersion:
\begin{equation}
  \alpha (z) = \left\{ \begin{array}{ll} -1 \quad & z<-1, \\
   z \quad & -1< z< 1, \\ 1 & z>1,  \end{array} \right. \label{dispform}
\end{equation}
which takes into regard saturation of the dispersion after
it has changed sign. This particular configuration can be
easily realized in experiments with DDF and, generally, adequately
represents the situation that we aim to consider.

We will consider the evolution (for $z>-1$) of a pulse represented by the
following natural initial configuration,
\begin{equation}
  u(z=-1,t) = A \sech (ht). \label{icform}
\end{equation}
Simulations demonstrate that
the evolution of other smooth, {\em unchirped} initial pulses is very similar
to that which is studied in detail below for the initial condition 
\eqref{icform}.

\section{Zero third-order dispersion}

If TOD is negligible, Eq. \eqref{mnls}
takes a simpler form,
\begin{equation}
iu_z + \alpha(z) u_{tt} + 2
|u|^2u = 0.\label{nls}
\end{equation}
Both equations \eqref{mnls} and \eqref{nls} conserve 
two quantities, viz., the ``optical energy''
\begin{equation}
  E = \int_{-\infty}^\infty |u(z,t)|^2 dt, \label{E}
\end{equation}
and the momentum
\begin{equation}
  P = \frac{i}{2} \int_{-\infty}^\infty (uu^*_t - u^*u_t )dt.
\end{equation}
These equations admit a Lagrangian representation. In particular, for 
the simplified equation \eqref{nls}, the {\it Lagrangian density} is
\begin{equation}
{\cal L} = \frac{i}{2}(u^*u_z-uu^*_z) - \alpha(z)|u_t|^2 + |u|^4 .
\end{equation}

Thus for zero TOD and $\alpha(z)$ chosen as per Eqs.
\eqref{dispform} and \eqref{icform}, the pulse evolution is governed
by the equations 
\begin{subequations}
  \label{vnls}
  \begin{align}
  & iu_z + z u_{tt} + 2|u|^2 u = 0, \qquad (-1<z<1) \label{vnlsa}\\
  & iu_z + u_{tt} + 2|u|^2 u = 0, \qquad (z>1) \,,\label{vnlsb}
\intertext{with}
  & u(-1,t) = A \sech (ht).
  \end{align}
\end{subequations}
The evolution can be characterized by the two parameters, amplitude $A$
and inverse size $h$ of the initial pulse \eqref{icform}, or, alternatively, 
by its conserved energy and initial ``mass''. The ``mass'', 
which is {\em not} a conserved quantity of the NLS equation, is defined as
\begin{equation}
  M = \int_{-\infty}^\infty |u(z,t)|\ dt.
\end{equation}
For the initial condition
\eqref{icform}, $M_0 \equiv M(z=-1) = \pi A/h$, and $E = 2A^2/h$.

Since dispersion is constant for $z>1$, the system \eqref{vnls}
can be considered as an
initial-value problem for the focussing NLS equation, with the initial
condition at $z=1$. Thus the
asymptotic solution will consist of a finite number of solitons represented
by the discrete spectrum, and
dispersive radiation represented by the corresponding continuous 
spectrum. The number of solitons 
can be found via the Inverse Scattering Transform for the NLS
equation by solving the Zakharov-Shabat (ZS) eigenvalue problem. 
To this aim, we define
\begin{equation} q(t) = u(z=1,t). \end{equation}
Assuming that $|q|$ decays as $|t|\to\infty$,
the ZS eigenvalue problem is based on the linear equations
\begin{subequations}
  \label{zs}
  \begin{align}
  & iv_t - \lambda v = q w, \\
  & iw_t + \lambda w = \bar q v
  \end{align}
\end{subequations}
for auxiliary {\it Jost functions} $u$ and $v$,
the overbar designating the complex conjugate.

The spectrum of the eigenvalue $\lambda$ consists of 
its continuous part on the real axis
and discrete eigenvalues, for which nontrivial solutions exist 
with functions $v$ and $w$ decaying exponentially as $|t|\to \infty$. 
The number
of the discrete eigenvalues in the upper half-plane is then equal to the 
number of 
solitons that will evolve from the wave packet $q(t)$. These 
discrete eigenvalues, in general, have a nonzero imaginary part. 

Note that, for a symmetric
$q$, which is the case here, if $\lambda$ is an eigenvalue, 
so also is $-\bar\lambda$. Solitary purely imaginary eigenvalues then
correspond to a single zero-velocity NLS soliton, while multiple
ones give rise to stationary {\it breathers}. If $\lambda$ has
a nonzero real part, then the pair $(\lambda,-\bar\lambda)$ 
corresponds to a pair of counter-propagating solitons 
having velocities of equal magnitude but opposite sign.

Thus, a two-stage process can be
used to solve Eqs. \eqref{vnls}. Firstly, Eq. \eqref{vnlsa}
is integrated from $z=-1$ to $z=1$, using a standard pseudo-spectral in $t$
and fourth-order in $z$ numerical method to obtain $q(t)$ at $z=1$. 
This is then
used in Eqs. \eqref{zs} to obtain the characteristics of the discrete
spectrum from the ZS eigenvalue problem. Particularly, we want to know
the total number of the discrete eigenvalues and the number of discrete
eigenvalues on the imaginary axis. These two
numbers can be found, using the methods elaborated in Refs. \cite{lewis85} 
and \cite{bronski96}. Introducing
\begin{equation} 
  \hat v = v\exp (-i\lambda t), \quad  \hat w = w\exp (i\lambda t),
\end{equation}
we integrate Eqs. \eqref{zs} over a finite region $t=[-L,L]$, where
$L \gg 1$, using the initial condition $\hat v (-L) = 1$ and finding
the {\it transmission coefficient}
$\hat v(L) = a(\lambda)$. The discrete eigenvalues are
then zeros of $a(\lambda)$ in the 
upper (${\rm Im} \{\lambda\} \geq 0$) complex half-plane. 
Since $a$ is an analytical function of
$\lambda$, the number of the eigenvalues
contained inside a closed curve $\Gamma$ on the complex plane is
\begin{equation}
 N (\Gamma) = \frac{1}{2\pi i} \int_{\Gamma} \frac{a'(\lambda)}{a(\lambda)}
 d\lambda.
\end{equation}

To obtain the total number of eigenvalues, let $\Gamma$ be the curve consisting
of the real axis and a semi-circle with an infinitely large radius in the upper
half-plane. On this infinite semi-circle, it can be
shown that $a(\lambda)\to1$, thus the
total number of eigenvalues is
\begin{equation}
 N = \frac{1}{2\pi i} \int_{-\infty}^{\infty}
  \frac{a'(\lambda)}{a(\lambda)}
 d\lambda.
\end{equation}

To obtain the number of discrete eigenvalues lying
exactly on the imaginary axis, let
$\Gamma_{{\rm im}}$ be the perimeter of 
a vanishingly thin but infinitely long rectangle enclosing
the positive imaginary axis. Then the total number of the 
static (zero-velocity) solitons is $N_{{\rm st}} 
= N(\Gamma_{\rm im})$.

Before proceeding to full numerical solutions, we can consider 
two different analytical
approximations for a solution of Eq. \eqref{vnlsa} in the interval
$z=[-1,1]$, valid, respectively, for long small-amplitude and short
large-amplitude pulses. For the long 
pulses, we define the variables
\begin{equation} x = ht, \quad u = A\psi. \end{equation}
Then \eqref{vnlsa} becomes
\begin{subequations}
  \label{psieq}
  \begin{align}
  & i\psi_z + z h^2\psi_{xx} + 2A^2|\psi|^2 \psi = 0, \\
\intertext{with}
  & \psi(z=-1,x) = \sech x,
  \end{align}
\end{subequations}
Because we now want to consider a long small-amplitude pulse, 
we assume that $h,A \ll 1$, such that
\begin{equation}
  h^2 = \mu, \quad A^2 = \gamma^2 \mu,
\end{equation}
where $\gamma$ is an $O(1)$ parameter and $\mu \ll 1$. We seek
a perturbation solution in the form
\begin{equation}
\psi = (R_0(x) + \mu^2 R_2 (z,x)+\dots) \exp \left[ 
i\mu(\phi_1(z,x)+\mu^2\phi_3(z,x)
+ \dots)\right] .
\label{pertsoln}
\end{equation}
Note that the $O(\mu)$ corrections, $R_1$ and $\phi_2$, can be shown to be 
identically zero. The initial condition (16b) then gives
\begin{equation}
R_0 = \sech (x).
\end{equation}
At $O(\mu)$ we obtain
\begin{equation}
R_0 \phi_{1z} = zR_{0xx} + 2\gamma^2 R_0^3,
\end{equation}
and therefore
\begin{equation}
\phi_1 = \half (z^2-1)(1-2\sech^2 x)+2\gamma^2(z+1)\sech^2 x.
\end{equation}
Next, at $O(\mu^2)$ it is found that $R_2$ satisfies the equation
\begin{equation}
R_{2z} = -z(2R_{0x}\phi_{1x} + R_0\phi_{1xx}),
\end{equation}
and consequently
\begin{multline}
R_2 =  2\sech^3 x(4-5\sech^2 x) \\ \times
\left(\frac{1}{4} (z^4-1)-\frac{1}{2} (z^2-1)-2\gamma^2 
\left( \frac{1}{3} (z^3+1)+\frac{1}{2} (z^2-1) \right) \right).
\end{multline}

\begin{figure}
\begin{center}
\includegraphics[width=12cm]{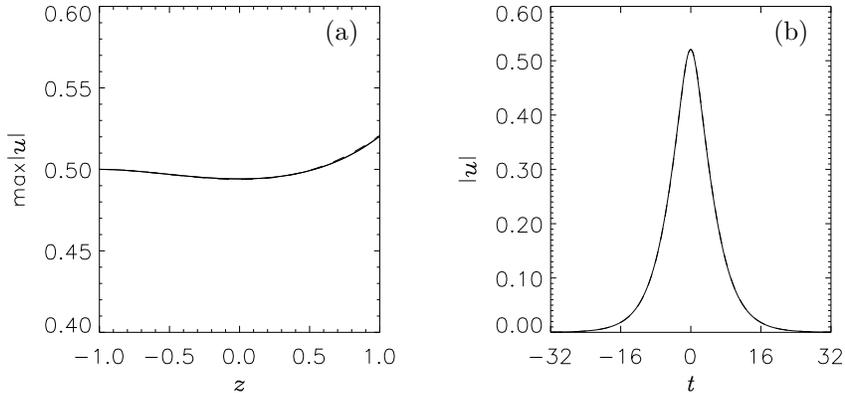}
\put(-30,145){\makebox(0,0){(b)}}
\put(-200,145){\makebox(0,0){(a)}}
\end{center}
\caption{A comparison of a numerical solution to Eq. \eqref{vnls} with the
perturbation solution \eqref{pertsoln} for $M_0=2\pi$ and $E=2$, which
correspond to 
$\mu=1/16$ and $\gamma=2$ in Eqs. (15) and (17). The numerical solution 
and perturbation solutions are shown, respectively, by solid and
dashed lines (that practically completely coincide).
(a) The field $|u|$ at $t=0$, and (b) $|u|$
at $z=1$.}
\label{fig1}
\end{figure}

The amplitude obtained from this perturbation solution
is compared with a numerical solution
in Fig. \ref{fig1}. As it is seen, the agreement is excellent.
This perturbation solution would be expected to be valid for
\begin{equation}
h^2 A^2 \ll 1 \txt{or} E \ll 2 (M_0/\pi)^\frac{3}{2}.
\end{equation}

In the limit $E\to 0$, while $M_0$ remains finite, it is apparent
that at $z=1$ the perturbation solution obtained by means of the
above method is
\begin{equation}
  q(t) \approx A\sech (ht),
\end{equation}
in which case
the variable-dispersion region is simply too short to have any effect on the
very long pulse. For this initial condition, a well-known exact solution 
to the ZS eigenvalue problem was found by Satsuma and Yajima \cite{satsuma74}. 
They showed that the solution consisted of $N$ stationary
solitons and a radiative component, where
\begin{equation}
  N = \left[ \frac{ 2M_0 + \pi }{\pi} \right],
\end{equation}
$[\cdot ]$ standing for the integer part. For $M_0 = N\pi$, where $N$ is 
a positive integer, the radiation component is absent, and the
solution consists purely of $N$ interacting solitons. Examples of such
solutions for $M_0 \gg \pi$ can be found in Ref. \cite{miller98}.

For short pulses, the variational approach proposed in Ref.
\cite{malo93} can be
used. To this end, we assume an {\em ansatz} for the wave packet,
\begin{equation}
  u(z,t) = A(z) \sech [t/a(z)] \exp \left[ i(\phi(z)+b(z)t^2)\right],
  \label{ans}
\end{equation}
where $A$, $a$, $\phi$ and $b$ are slowly varying real functions of $z$. 
Varying the averaged Lagrangian,
\begin{equation}
  L(z; A,\phi, b, a) = \int_{-\infty}^{\infty} {\cal L}\ dt,
\end{equation} with respect to these
free parameters, it can be shown that $A$, $\phi$ and $b$ can
be eliminated in favor of the pulse width, $a$, as
\begin{subequations}
  \label{veqns}
  \begin{align}
  & \frac{dE}{dz} = \frac{d}{dz} (2aA^2) = 0, \\
  & b = \frac{1}{4\alpha a}\frac{da}{dz}, \\
  & \frac{d\phi}{dz} = -\frac{2\alpha}{3a^2} + \frac{5E}{6a},
  \end{align}
\end{subequations}
where $\alpha(z)$ is the variable dispersion coefficient in Eq.
\eqref{vnlsa} ($\alpha(z)=z$ in the case under consideration),
and $a(z)$ satisfies another variational equation,
\begin{equation}
\frac{d}{dz} \left( \frac{1}{\alpha}\frac{da}{dz} \right)
    + \left( \frac{4}{\pi} \right)^2 \left( 
    \frac{E}{2a^2} - \frac{\alpha}{a^3}\right) = 0. \label{veqn}
\end{equation}
Since the initial chirp is zero, the appropriate initial conditions for
\eqref{veqn} are
\begin{equation}
  a = h^{-1}, \quad \frac{da}{dz} = 0 \txt{at} z = -1. \label{vic}
\end{equation}

\begin{figure}
\begin{center}
\includegraphics[width=12cm]{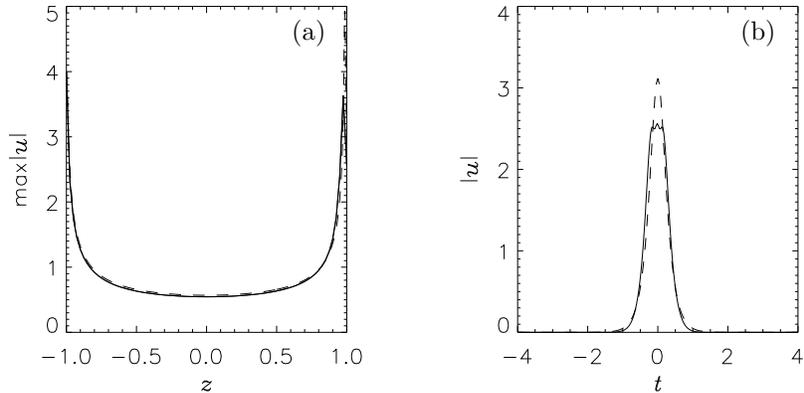}
\put(-30,145){\makebox(0,0){(b)}}
\put(-200,145){\makebox(0,0){(a)}}
\end{center}
\caption{A comparison of a numerical solution of Eq. \eqref{vnls} with the
variational approximation \eqref{ans} for $M_0=\pi/2$ and $E=4$. The numerical 
and variational solutions are shown by solid and dashed lines, respectively.
(a) The field $|u|$ at $t=0$, and (b) $|u|$
at $z=1$.}
\label{fig2}
\end{figure}

In Fig. \ref{fig2}, an example of a numerical solution to Eqs. 
\eqref{veqn}--\eqref{vic}
is compared with a full numerical solution of \eqref{vnls}. The agreement
between the two solutions is quite good, except for a narrow
region near the maximum of $|u(t)|$
at $z=1$. Subsequently, the solutions experience rapid variations with
$z$ which the variational ansatz is unable to capture.

\begin{figure}
\begin{center}
\includegraphics[width=9cm]{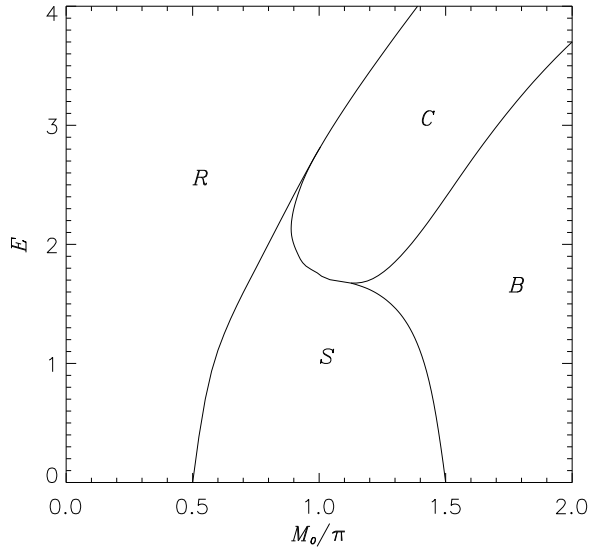}
\end{center}
\caption{A classification of numerical solutions to Eq. \eqref{vnls} in terms
of the energy $E$ and initial mass $M_0$. 
$R$ denotes the region of the radiative solutions, 
$S$ the single-soliton solutions, $B$ the bound-state (breather)
solutions, and $C$ the pair of counter-propagating solitons.}
\label{ClassFig}
\end{figure}

Using these analytical approximations and direct numerical solutions,
a few different types of the solutions to Eq. \eqref{vnls} can be 
identified, varying the initial
mass and energy of the pulse. The results are summarized
in Fig. \ref{ClassFig}. In the range considered, four particular 
types of the solutions were found. To construct this diagram, many numerical
simulations were required, with particular attention being taken to
determine the boundaries between the different regions. Of course,
these boundaries can only be determined to the accuracy shown here by
using the two-stage numerical process described above. If only
direct numerical simulations are used, it becomes very difficult to gauge
when a transition occurs. For example, a prohibitively large integration time
is needed to distinguish between the decaying-radiation and 
small-amplitude solitons, 
whereas this distinction can easily be made by considering
the ZS eigenvalue problem.

\begin{figure}
\begin{center}
\includegraphics[width=11cm]{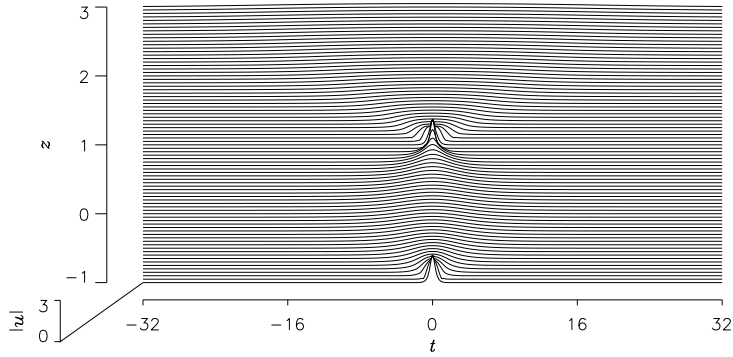}
\end{center}
\caption{The evolution of $|u|$ for 
a radiative (no-soliton) solution of Eq. \eqref{vnls} with $E=2$ and 
$M_0=\pi/2$.}
\label{radsoln}
\end{figure}

For a small initial mass $M_0$, we find no discrete eigenvalues
of the ZS scattering problem, so the pulse completely decays into 
dispersive radiation, in which case $|u|$ has a maximum at $t=0$ and is
monotonically decaying with $|t|$. It is worth noting that 
near the $R$--$C$ boundary, 
a double-humped structure can form at $z>1$ with symmetric
local maxima away from $t=0$ (which resembles a structure found recently
in the same model with the reverse direction of the dispersion 
change \cite{ccgm99}); however, eventually the pulse
decays into a monotonically decreasing structure. An example of such
a solution in shown in Fig. \ref{radsoln}. For $M_0 \to 0$, the evolution
of the pulse at $z>1$ is reasonably well approximated by the solution
of the variational equation \eqref{veqn}. 

\begin{figure}
\begin{center}
\includegraphics[width=11cm]{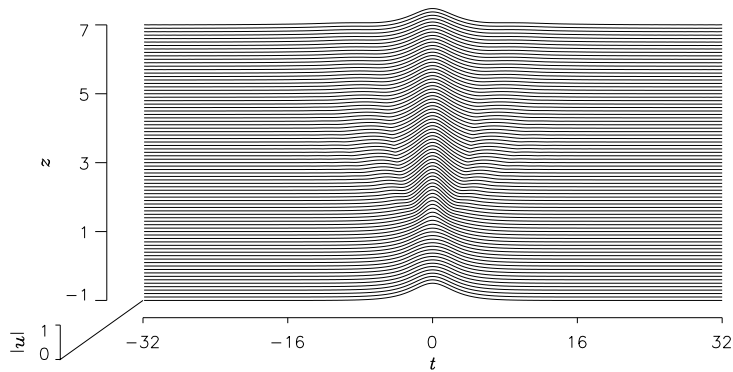}
\end{center}
\caption{The evolution of $|u|$ for 
a single-soliton solution of Eq. \eqref{vnls} with $E=1$ and $M_0=\pi$.
Here, approximately 95\% of the energy of the initial pulse is 
transmitted to the soliton.}
\label{solsoln}
\end{figure}

As $M_0$ is increased,
two possibilities occur. In the first case, which occurs for small values
of $E$, a single discrete eigenvalue appears on the imaginary axis.
This corresponds to the ``mass'' of the pulse now being sufficiently large
to form a soliton. Thus the asymptotic solution consists of a stationary
soliton and dispersive radiation, see the example in Fig. \ref{solsoln}.

\begin{figure}
\begin{center}
\includegraphics[width=11cm]{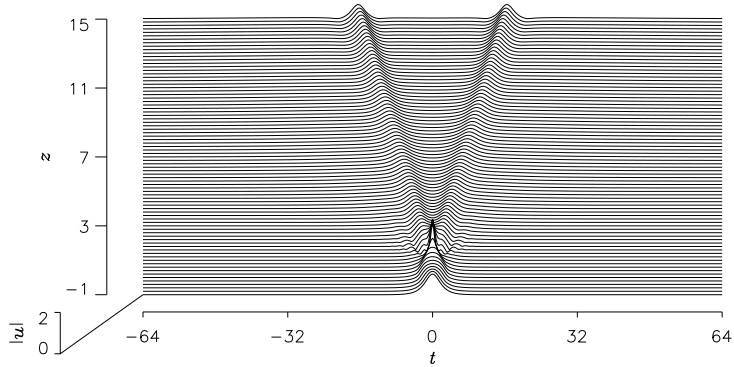}
\end{center}
\caption{The evolution of $|u|$ for a pair of counter-propagating solitons
in Eq. \eqref{vnls} with $E=3$ and $M_0=3\pi/2$.
Here, approximately 93\% of the energy of the initial pulse is 
transmitted to the two solitons.}
\label{countsoln}
\end{figure}

The second case occurs for larger values of $E$. Here, for sufficiently
large values of the initial mass, the chirp has reduced sufficiently
for the above-mentioned double-humped structure
to develop into a pair of {\em counter-propagating} solitons. This corresponds
to a pair of complex discrete eigenvalues appearing out of the real
axis, where $\lambda = \pm \lambda_{{\rm r}}
+i\lambda_{{\rm i}}$, $\lambda_{{\rm r}}$ and $\lambda_{{\rm i}}$ being
both real. An example is shown in Fig. \ref{countsoln}. 

\begin{figure}
\begin{center}
\includegraphics[width=11cm]{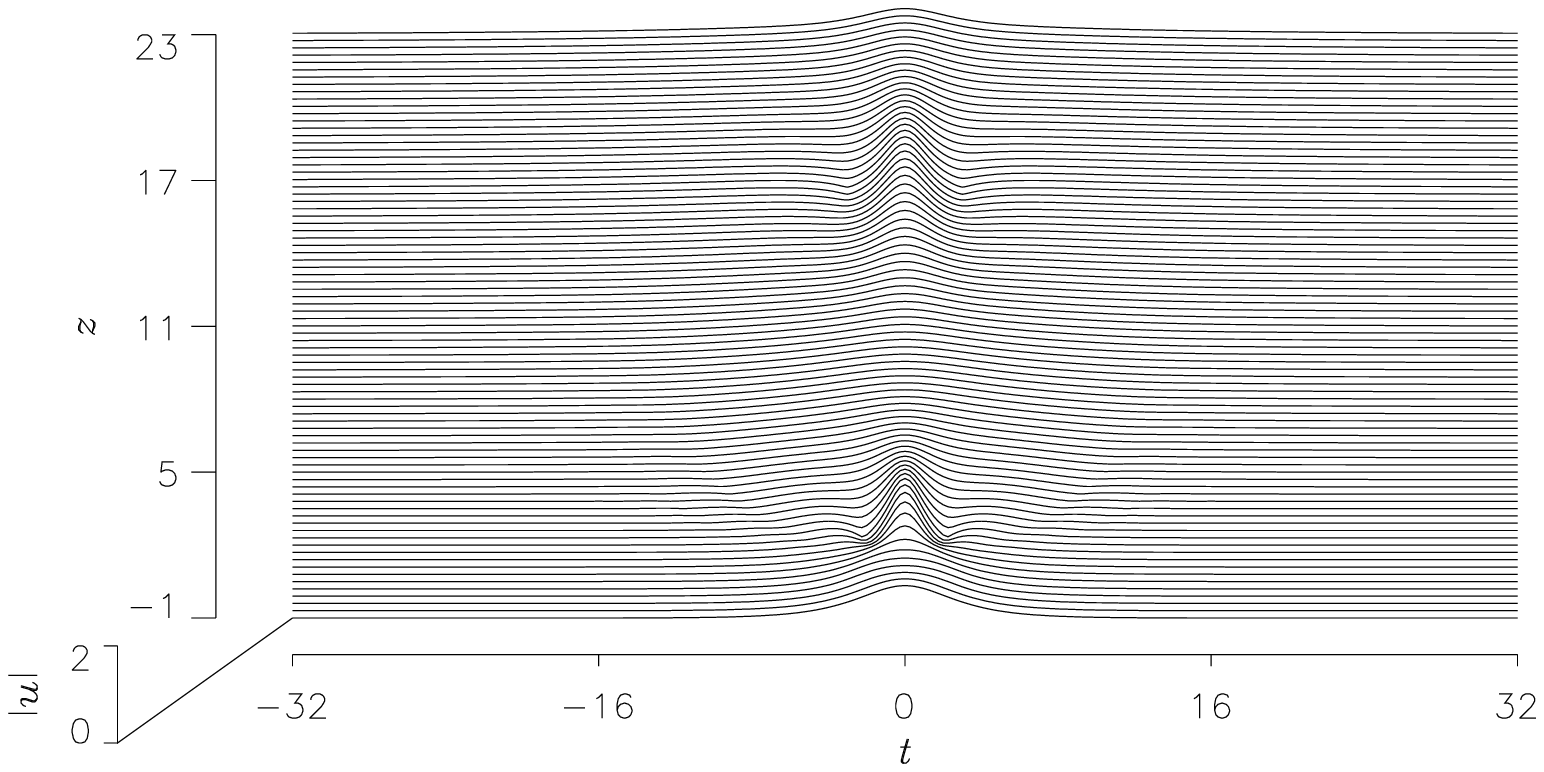}
\end{center}
\caption{The evolution of $|u|$ for a bound-state (breather) solution
of Eq. \eqref{vnls} with $E=2$ and $M_0=3\pi/2$.
Here, between 90--95\% of the energy of the initial pulse is 
transmitted to the breather.}
\label{boundsoln}
\end{figure}

In the last regime, a stationary {\it breather},
or a bound-state solution forms which asymptotically consists of a pair
of interacting solitons. An example of this is shown in Fig. 
\ref{boundsoln}. This can happen in two different ways.
If the $C\to B$ transition is considered, this occurs when the chirp 
has reduced sufficiently to no longer cause the splitting of the solitons.
In terms of the ZS eigenvalue problem, this corresponds to the two eigenvalues
approaching each other at a point on the imaginary axis. After their
``collision" on the imaginary axis,
two purely imaginary eigenvalues appear, corresponding to two
zero-velocity solitons. For the $S\to B$
 transition, this simply corresponds to the
``mass'' having increased sufficiently to give rise to a new zero-velocity
soliton. This second
soliton, like the one that forms in the $R\to S$ transition,
appears out of the origin and corresponds to a purely imaginary eigenvalue.

Within the framework of the ZS eigenvalue problem, generic transitions 
that lead to the formation
of a pair of counter-propagating solitons take place when two eigenvalues appear
out of the continuous spectrum, such as occurs for the $R\to C$ transition, or
two imaginary eigenvalues collide and subsequently reshape into a
pair of complex eigenvalues, which occurs for the $B\to C$
transition. No evidence has previously been reported of a transition where
a single imaginary eigenvalue splits into a pair of complex 
eigenvalues. Thus, from theoretical considerations, we would expect a $B$
region to always occur between the $S$ and $C$ ones, giving rise to a transition
chain
$S\to B\to C$. This intermediate $B$ layer must disappear at 
a {\it triple point}, where the $R$ and $C$
regions become adjacent, as again the transition $R\to C$ is admitted
by the general analysis. For $M_0 > 1.1$ the $S\to B\to C$
transition occurs as expected; however for $M_0 < 1.1$ no evidence of the
above-mentioned intermediate $B$ layer could be found. Nevertheless, because 
a direct splitting of a soliton into a pair of symmetric 
counterpropagating ones obviously contradicts the principle of
continuity and therefore cannot take place in a model governed by a 
smooth differential equation, we conjecture that the missing intermediate 
$B$ layer does exist, but it is too thin to be detected using the 
current numerical techniques.

If the initial ``mass'' of the pulse is fixed, say, to the value $M_0=1$, it is
apparent that as the energy is increased the number of solitons goes through
the transition $1\to2\to0$. This result appears counter-intuitive, as one would expect the number of solitons to monotonically
increase with the energy. However, this
transition can be explained by the influence of the pulse's chirp at $z=1$. 
As $E$ is increased,
the strength of the chirp increases too. Thus, while for small values of $E$ 
the chirp may have little
effect on the pulse and one soliton is able to form, for larger values 
of the energy the chirp becomes strong enough to split the single soliton into
 a pair of
solitons \cite{Egypt}. For even larger values of the chirp, this pair 
is destroyed.
A possibility could be to split the two solitons into four moving ones;
however, because the mass at $z=1$ is not sufficient to form four solitons,
the pulse decays into radiation.

\section{Finite third-order dispersion}

In this section, we consider solutions to Eq. \eqref{mnls}, supplemented by Eqs.
\eqref{dispform} and \eqref{icform}, under the assumption that $\delta$
is small
($\delta \ll 1$) but finite. Therefore, the pulse evolution is now governed
by the system
\begin{subequations}
  \label{vmnls}
  \begin{align}
  & iu_z + z u_{tt} + 2|u|^2 u = i\delta u_{ttt}, \qquad (-1<z<1) \\
  & iu_z + u_{tt} + 2|u|^2 u = i\delta u_{ttt}, \qquad (z>1) \\
\intertext{with}
  & u(z=-1,t) = A \sech (ht). 
  \end{align}
\end{subequations}
For $z>1$, we are again dealing with a constant-coefficient 
equation, which is, however, no longer integrable (note that
this equation, with both constant and periodically
modulated coefficient in front of the usual 
dispersion term, has attracted a lot of attention in nonlinear
fiber optics, see, e.g., Refs. \cite{desa90,fran98,labo99} and references
therein). Thus, only direct 
numerical simulations can be used to obtain the solutions.

In the consideration of the transition from the anomalous to normal
dispersion (i.e., focussing to defocussing), 
which was the subject of Ref. \cite{ccgm99}, the pulse width
achieved a minimum at the zero-dispersion point $z=0$, and consequently
TOD was becoming more important as this point was approached. However,
in the present case the
pulse width achieves a {\em maximum} at $z=0$ and, in general, a minimum
near $z=1$. Thus, due to the the underlying assumption $|\delta| \ll 1$,
it would not be expected that TOD would be significant until well into the
focussing (anomalous-dispersion) region. Hence the effect of 
TOD on the pulse can be estimated by
comparing the magnitude of the two constant-coefficient dispersion terms.
If the pulse width is $O(a)$, then 
\begin{equation}
  \left| \frac{\delta u_{ttt}}{u_{tt}} \right| \sim \frac{|\delta|}{a}.
\end{equation}
From Figs. \eqref{radsoln}--\eqref{boundsoln} it is apparent that, for all 
the cases, $a_{\min} < h^{-1}$, therefore TOD can be neglected if
\begin{equation}
  |\delta| \ll \frac{2M_0^2}{\pi^2 E}.
\end{equation}

\begin{figure}
\begin{center}
\includegraphics[width=11cm]{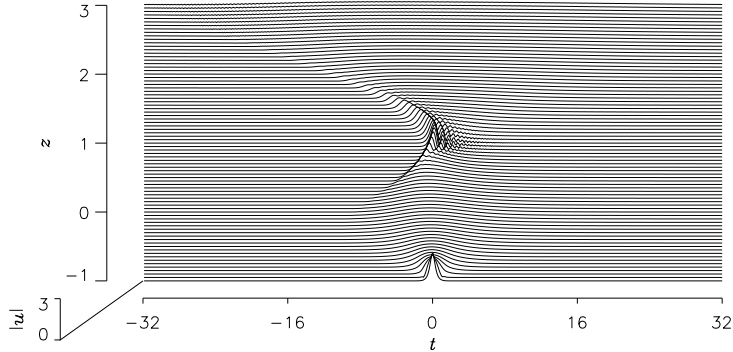}
\end{center}
\caption{The evolution of $|u|$ for the solution of Eq. \eqref{vmnls}
with $E=2$, $M_0=\pi/2$ and $\delta=0.01$, demonstrating the effect of
the third-order dispersion on the radiative solution which was shown,
for the $\delta =0$ case, in Fig. \ref{radsoln}.}
\label{fig:7}
\end{figure}

When TOD is taken into regard, it manifests itself in three ways. 
Firstly, TOD destroys the symmetry of the pulse. This is most dramatically
seen if one considers the effect of TOD on the radiative solutions, an
example of which is shown in Fig. \ref{fig:7} for a small
value of the TOD coefficient, $\delta =0.01$. 
In the defocussing (normal-dispersion) region ($-1<z<0$), 
it is apparent that TOD has little effect.
However, in the focussing (anomalous-dispersion) region $(z>0)$, 
it is apparent that TOD completely 
destroys the symmetry of the pulse well before the minimum of
the pulse width is reached. 
Comparing this with Fig. \ref{radsoln}, it is apparent that,
nevertheless, the
effect of TOD far away from the zero-dispersion point
does not change the radiative type of the solution.

The second effect of TOD, which is now well known,
is that solitons are no longer localized, but rather generate
a small-amplitude co-propagating oscillatory tail (such
nonlocal solitary waves have been termed ``nanopterons'' by Boyd
\citep{boyd98}). Resonant generation of the tails in the 
context of the modified NLS equation including the TOD term
has been considered in several works, see, e.g., Refs. \cite{wai90,grim95}.
For a single soliton with speed $c$ and amplitude $A$, it was
demonstrated in Ref. \cite{grim95} that, for $\delta \ll 1$,
the tail has a wavenumber
\begin{equation}
  k_r = -\delta^{-1} -  c + O(\delta), \label{krgrim}
\end{equation}
and amplitude
\begin{equation}
  A_r \approx \frac{\pi K}{\delta} \exp \left( -\frac{\pi}{2\delta A} \right),
    \label{argrim}
\end{equation}
where $K \approx 8.58$. For the single-soliton solutions considered
in section 2, the tail generation is a major consequence of TOD (not shown here).
  
\begin{figure}
\begin{center}
\includegraphics[width=11cm]{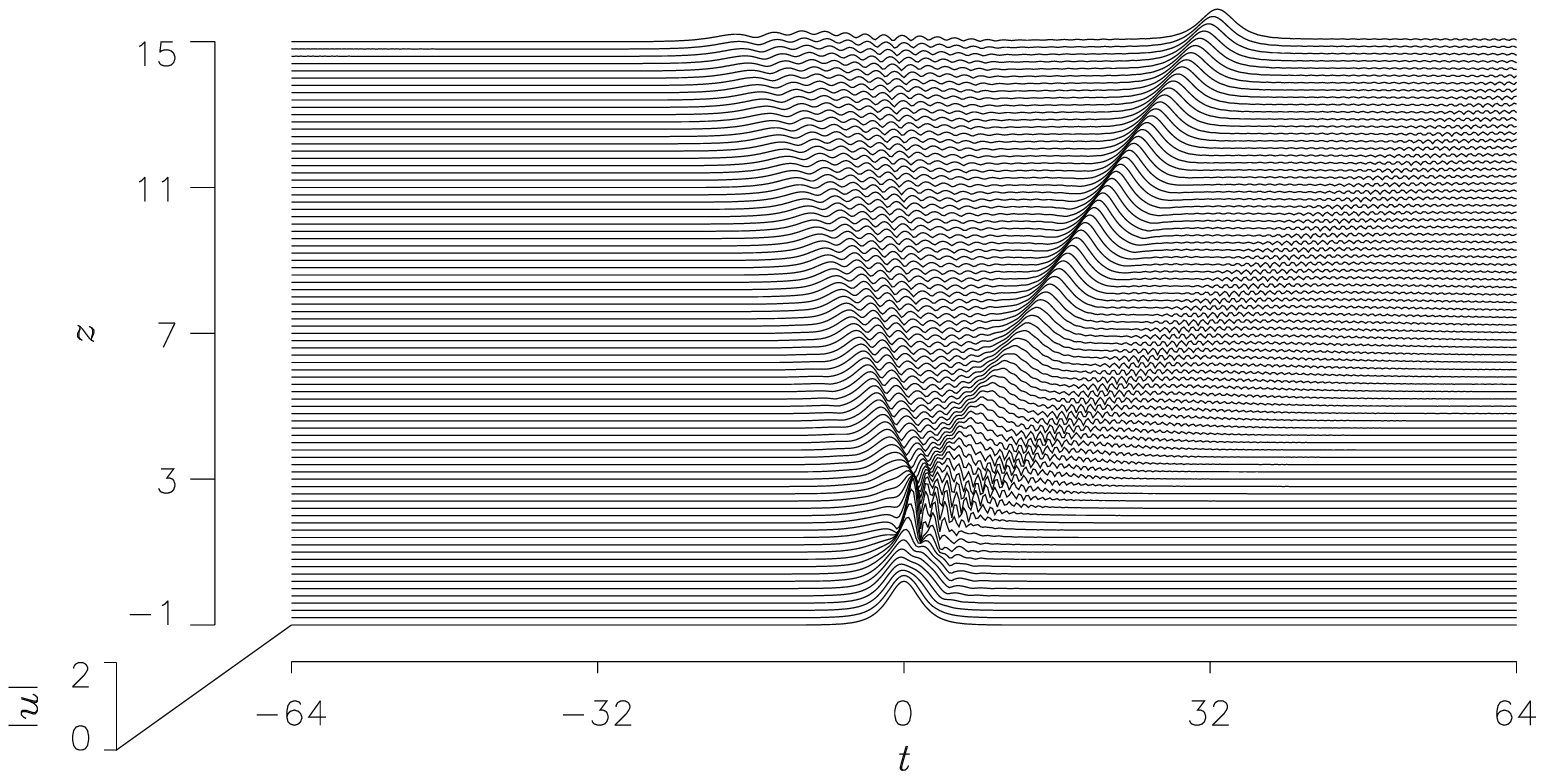}
\put(-40,170){\makebox(0,0){(a)}}
\end{center}
\begin{center}
\includegraphics[width=9.5cm]{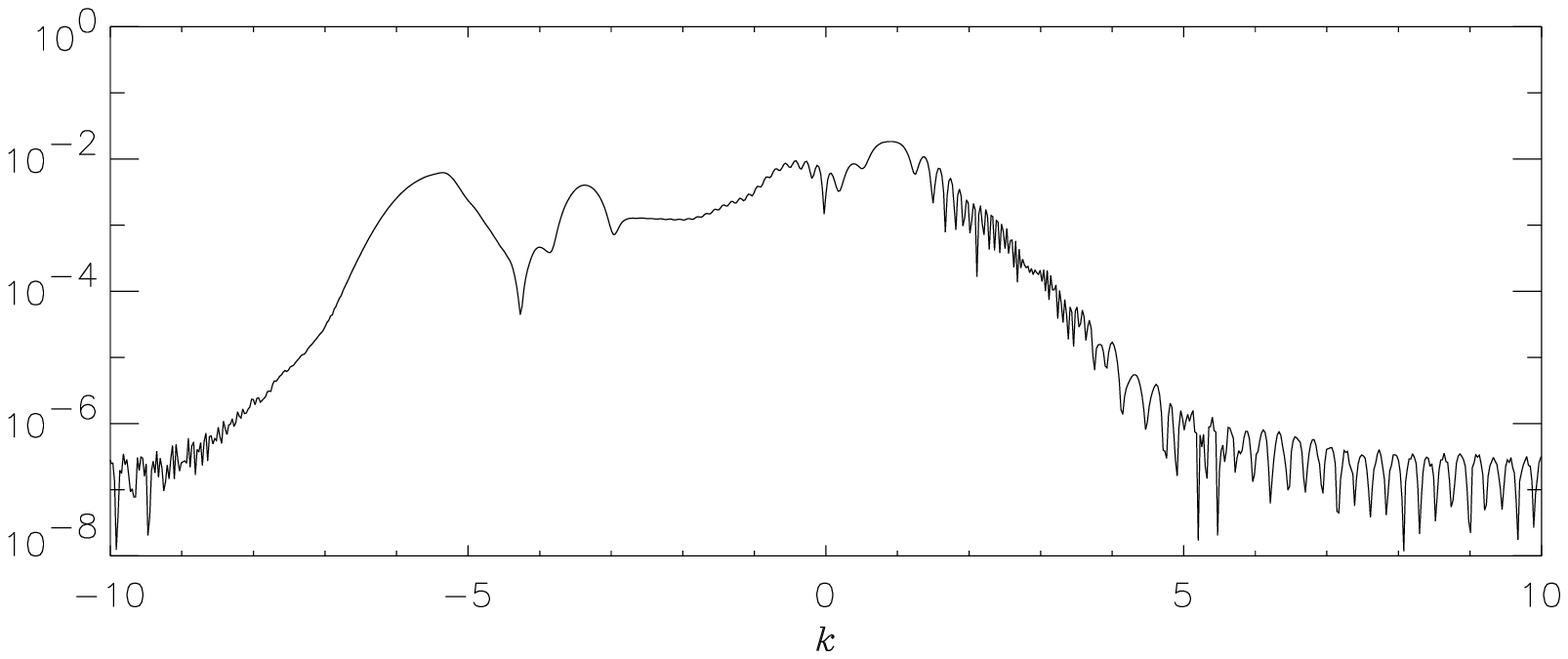}
\put(-30,95){\makebox(0,0){(b)}}
\end{center}
\caption{The solution of Eq. \eqref{vmnls}
with $E=3$, $M_0=3\pi/2$ and $\delta=0.2$, demonstrating the effect of the
third-order dispersion on the counter-propagating soliton-pair solution, 
which was shown for case $\delta =0$ in Fig. \ref{countsoln}.
(a) The evolution
of the field $|u|$, (b) the 
absolute value of the Fourier transform at $z=15$
(note that the vertical scale in the panel (b) is logarithmic).}
\label{fig:8}
\end{figure}

The symmetry breaking and appearance of the resonant oscillatory waves are
clearly apparent if the effect of TOD on the counter-propagating
soliton pair solutions is considered. An example is shown in Fig.
\ref{fig:8}. With $\delta = 0$, both solitons had the same amplitude and
equal but opposite speeds. However, nonzero $\delta$ destroys this
symmetry. For positive $\delta$, the leftward-propagating
soliton is now reduced in amplitude and its speed decreases, whereas
for the rightward-propagating soliton the amplitude and speed {\em increase}.
The co-propagating oscillatory waves (tail generation)
can be seen both in the evolution
of the wave field in Fig. \ref{fig:8}(a), and by examining its Fourier transform,
which is shown (on the logarithmic scale)
in Fig. \ref{fig:8}(b) at $z=15$. For the smaller
soliton, the tail can be observed to the right
of it, corresponding to the peak in the Fourier transform
at $k\approx-3.5$. For the larger soliton, it is more difficult to
observe its tail in the evolution plot,
however the front of this wave group can be seen starting to propagate
rightwards
from the pulse at $z\approx 2$. These waves correspond
to the well-pronounced peak in the Fourier transform at $k\approx -5.5$

\begin{figure}
\begin{center}
\includegraphics[width=11cm]{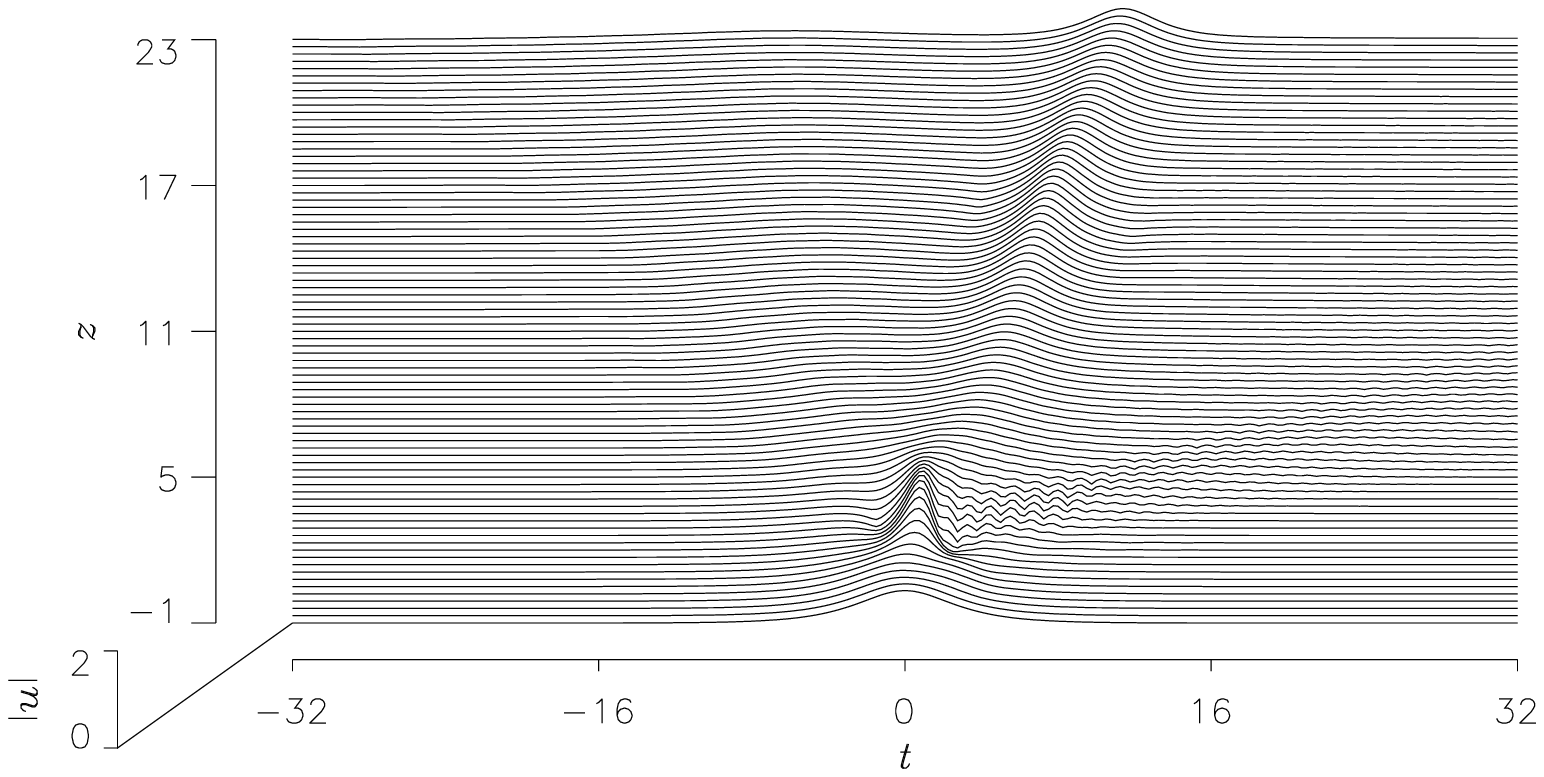}
\put(-40,170){\makebox(0,0){(a)}}
\end{center}
\begin{center}
\includegraphics[width=9.5cm]{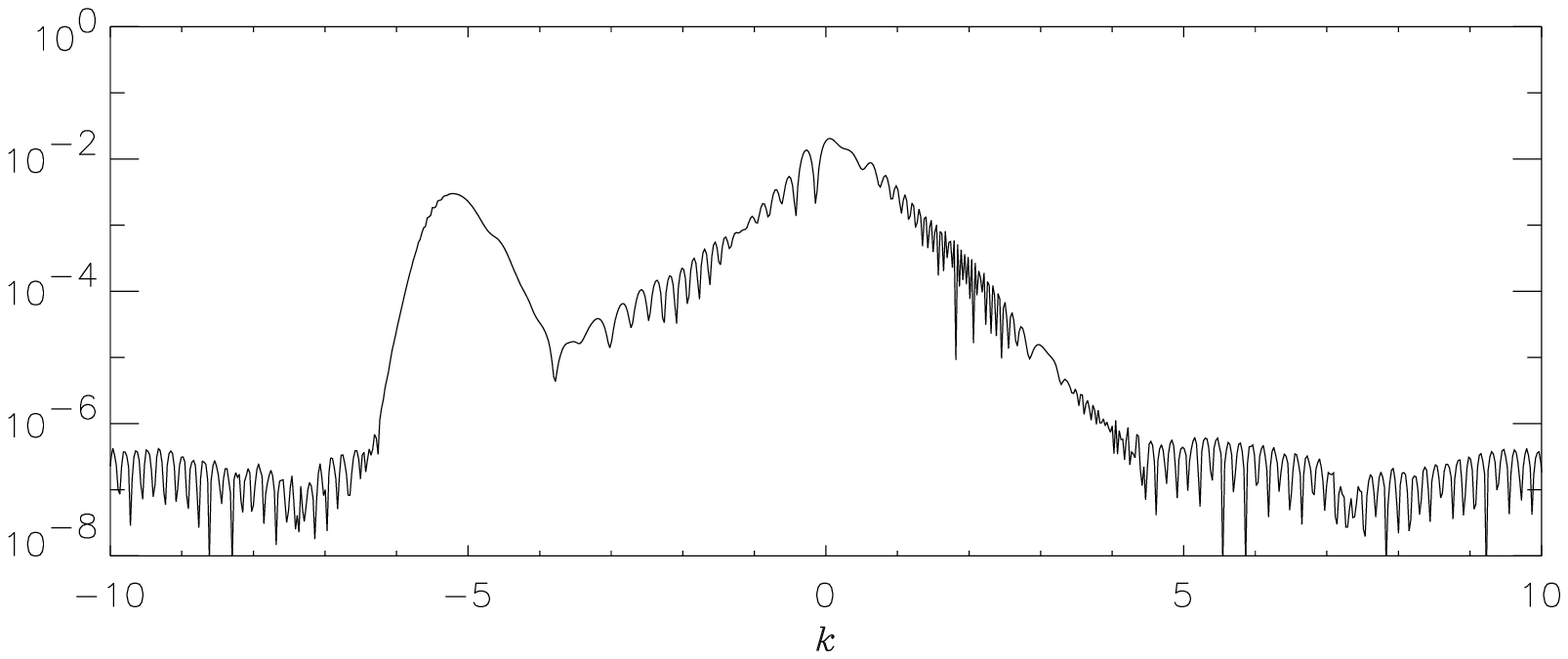}
\put(-30,95){\makebox(0,0){(b)}}
\end{center}
\caption{The solution of Eq. \eqref{vmnls}
with $E=2$, $M_0=3\pi/2$ and $\delta=0.2$, demonstrating the effect of
the third-order dispersion on the bound-state solution shown in 
Fig. \ref{boundsoln}.
(a) The evolution
of the field $|u|$; (b) the 
absolute value of the Fourier transform at $z=23$.}
\label{fig:6}
\end{figure}

Finally, we consider the effect of TOD on the breather
that could exist at $\delta =0$. In this case, a major
effect is {\em splitting} of the breather into two
asymmetric solitons, provided that $\delta$ is sufficiently
large, an example of which is shown in Fig. \ref{fig:6}.
Note that a possibility of splitting of a pulse under the 
action of TOD is also known in the constant-coefficient
NLS equation, see, e.g., Ref. \cite{desa90}.

In the case shown in Fig. \ref{fig:6}, the main breather-like soliton 
propagates to the right, but there is
also a very weak wave propagating to the left. Whether the
leftward-propagating wave is a soliton or just a packet 
of dispersive radiation cannot
be determined. As in Fig. \ref{fig:7}, there are oscillatory tail
waves co-propagating with the main soliton. The front of these waves can
be seen propagating rapidly to the right, and the peak in the Fourier
transform at $k\approx-5.3$ (Fig. \ref{fig:6}(b)) is associated 
with these waves.
In simulations of the same case, but with $\delta=0.1$ (twice as small,
not shown here), the main breather soliton keeps a nearly zero velocity,
and no small-amplitude leftward-propagating wave was detected. 
Qualitatively, this latter case, with the smaller value of $\delta$, is very
similar to what was shown in Fig. \ref{boundsoln}.

For each of the large-amplitude waves in
Figs. \eqref{fig:8} and \eqref{fig:6}, a comparison of the 
observed co-propagating tail waves can be made
with the predictions of Ref. \cite{grim95}, namely the
expressions \eqref{krgrim} and
\eqref{argrim}. This is shown in Table \eqref{tab2}. 
For the small-amplitude soliton from Fig. \ref{fig:8}, the comparison
was not made, as the presence of the soliton with the larger amplitude
and its tail renders the asymptotic analysis invalid for the
smaller soliton. As can be seen from Table \ref{tab2} for the
dominant rightward propagating soliton, the agreement between the theoretically predicted
and observed values of the amplitude is good; however, there is
a conspicuous discrepancy, especially in the case shown
in Fig. \ref{fig:8}, in the values
of the wavenumber. Plausibly, the discrepancy can be fixed by
including the next-order term in the expansion \eqref{krgrim};
however, one must also calculate the phase of the solitons in 
order to do this.

\begin{table}
\begin{center}

\begin{tabular}{ c c c c c c c}

\noalign{\hrule}
\noalign{\bigskip}

Figure  & $A$ & $c$ & $k_r$ & $A_r$ & $k_o$ & $A_o$ \\

\noalign{\bigskip}
\noalign{\hrule}
\noalign{\bigskip}

\eqref{fig:8} & $0.8$ & $2.3$ & $-7.3$ & $7\times 10^{-3}$ & $-5.5$ & 
$6\times 10^{-3}$ \\
\noalign{\smallskip}
\eqref{fig:6} & $0.7$ & $0.45$ & $-5.45$ & $2\times 10^{-3}$ & $-5.3$ & 
$3\times 10^{-3}$ \\

\noalign{\bigskip}
\noalign{\hrule}

\end{tabular}
\end{center}

\caption{A summary of predicted and observed amplitudes and wavenumbers
for the oscillatory tail waves seen in Figs. \eqref{fig:8} and \eqref{fig:6}, 
where $\delta=0.2$. $A$
and $c$ are, respectively, the soliton amplitudes and speeds, $k_r$ and
$A_r$ are the wavenumbers and amplitudes predicted by 
Eqs. \eqref{krgrim} and \eqref{argrim}, while $k_o$ and $A_o$ are the
observed ones.}
\label{tab2}

\end{table}

\section{Conclusion}

In this work, we have analyzed in detail self-trapping of a soliton
from a wave packet that passes from a defocussing region into a
focussing one in a spatially inhomogeneous nonlinear waveguide,
described by a variable-dispersion NLS equation,
in which the dispersion coefficient changes its sign from normal
to anomalous. The model can be realized in terms of two (at least) 
very different but realistic physical applications: a dispersion-decreasing 
nonlinear optical fiber, and natural waveguides for internal waves in the 
ocean. It was found that, depending on the values of the
(conserved) energy and (nonconserved) ``mass"
of the initial pulse, four qualitatively different outcomes of
the pulse transformation are possible: decay into radiation; self-trapping
into a single soliton; formation of a breather; and formation of a
pair of counterpropagating solitons. A chart of the corresponding
parametric plane has been drawn, which demonstrates some
unexpected features. One of them is that, with the increase of the energy, 
while the initial ``mass" is kept constant, a soliton, a
pair of the counterpropagating solitons, or a breather 
eventually decay into pure radiation. Another noteworthy
feature is that a direct transition from
a single soliton to a pair of symmetric counterpropagating 
ones seems virtually
possible. An explanation for these features was proposed. In two
cases when analytical approximations apply, viz., a straightforward
perturbation theory for broad initial pulses, or the variational
approximation for narrow ones, comparison with the direct simulations
shows a good agreement. 

\section*{Acknowledgements}

B.A.M. appreciates a support from the Department of Mathematics and
Statistics at Monash University (Clayton, Australia) and from the
Australian Research Council grant No. A89927007.


\end{document}